\documentclass[fleqn,usenatbib]{mnras}

\usepackage{newtxtext,newtxmath}

\usepackage[T1]{fontenc}

\DeclareRobustCommand{\VAN}[3]{#2}
\let\VANthebibliography\thebibliography
\def\thebibliography{\DeclareRobustCommand{\VAN}[3]{##3}\VANthebibliography}

\usepackage{graphicx}	
\usepackage{amsmath}	

\usepackage{amssymb}	%

\usepackage[dvipsnames]{xcolor}

\newcommand\tE{t_{\rm E}}

\newcommand\tEnk{t_{{\rm E}_{n,k}}}

\newcommand{\edit}[1]{#1}
\newcommand{\newedit}[1]{#1}

\title[Spatially resolved microlensing timescales]{Spatially resolved microlensing timescale distributions across the Galactic bulge with the VVV survey}

\author[Z. Kaczmarek et al.]{Zofia Kaczmarek$^{1,2}$\thanks{E-mail: zofia.kaczmarek@uni-heidelberg.de}, Peter McGill$^{3}$, N. Wyn Evans$^{1}$, Leigh C. Smith$^{1}$, Nathan Golovich$^{3}$,
\newauthor Eamonn Kerins$^{4}$, David Specht$^{4}$,  William A. Dawson$^{3}$
\\
$^{1}$Institute of Astronomy, University of Cambridge, Madingley Rd, Cambridge CB3 0HA, UK \\
$^{2}$Zentrum f{\"u}r Astronomie der Universit{\"a}t Heidelberg, Astronomisches Rechen-Institut, M{\"o}nchhofstr. 12-14, 69120 Heidelberg, Germany \\
$^{3}$Space Science Institute, Lawrence Livermore National Laboratory, 7000 East Avenue, Livermore, CA 94550, USA \\
$^{4}$Jodrell Bank Centre for Astrophysics, Department of Physics and Astronomy, University of Manchester, Oxford Road, Manchester M13 9PL, UK
}

\date{Accepted XXX. Received YYY; in original form ZZZ}

\pubyear{2024}

\begin{document}
\label{firstpage}
\pagerange{\pageref{firstpage}--\pageref{lastpage}}
\maketitle

\begin{abstract}
We analyze 1602 microlensing events found in the VISTA Variables in the Via Lactea (VVV) near-infrared (NIR) survey data.
We obtain spatially-resolved, efficiency-corrected timescale distributions across the Galactic bulge ($|\ell|<10^\circ,$ $|b|<5^\circ$), using a Bayesian hierarchical model.
Spatially-resolved peaks and means of the timescale distributions, along with their marginal distributions in strips of longitude and latitude, are in agreement at a 1$\sigma$ level with predictions based on the Besançon model of the Galaxy. 
We find that the event timescales in the central bulge fields ($|\ell| < 5^\circ$) are on average shorter than the non-central ($|\ell| > 5^\circ$) fields\edit{, with the average peak of the lognormal timescale distribution at 23.6 $\pm$ 1.9 days for the central fields and 29.0 $\pm$ 3.0 days for the non-central fields.}
Our ability to probe the structure of the Bulge with this sample of NIR microlensing events is limited by the VVV survey's sparse cadence and relatively small number of detected microlensing events compared to dedicated optical surveys. Looking forward to future surveys, we investigate the capability of the {\it Roman} telescope to detect spatially-resolved asymmetries in the timescale distributions.
We propose two pairs of \textit{Roman} fields, centred on ($\ell = \pm 9,5^\circ$, \edit{$b=-0.125^\circ$}) and ($\ell = -5^\circ$, $b=\pm 1.375^\circ$) as good targets to measure the asymmetry in longitude and latitude, respectively.
\end{abstract}




\section{Introduction}

\cite{Griest1991bulge} and \cite{Paczynski1991} first proposed monitoring the Milky Way bulge for microlensing events, partly as a reliability check on the results towards the Magellanic Clouds \citep{Alard1995, Evans2003}. The early measurements of the photometric microlensing optical depth (or the probability that a microlensing event occurs) towards the bulge turned out to be higher than expected \citep{Udalski1994,Al95}. Exploiting this high rate, modern microlensing surveys of the Milky Way bulge provide the largest source of microlensing events. They detect about $2000$ photometric events per year \citep{Udalski2015,KMTnet2016}, which have proved useful for a range of different astrophysical applications \citep{Gould2001,Mao2012}. Most notably, short timescale deviations, caused by planets around the lens, have been detected for $>100$\footnote{\url{https://exoplanetarchive.ipac.caltech.edu/docs/counts_detail.html}} events to date \citep[e.g.,][]{Bond2004,Udalski2005,Beaulieu2006}. These events have provided a probe of low-mass planets (down to Earth-mass) at intermediate separations from their hosts ($\approx1$ au) which are typically off-limits to other planet detection techniques \citep{Tsapras18}.

For the majority of photometric events found via the monitoring channel, and in the absence of any higher-order effects, the only parameter that can be extracted containing physical information is the Einstein timescale \citep{Paczynski1986halo},
\begin{equation}
    t_{\rm E} = \frac{\theta_{\rm E}}{\mu_{\text{rel}}} =\frac{1}{v_{\text{rel}}}\sqrt{\frac{4 GM_{\rm L}}{c^2} \frac{D_{\rm L} (D_{\rm S}-D_{\rm L})}{D_{\rm S}}}.
    \label{eq:einstein_timescale_formula}
\end{equation}
This is the time taken for the source to traverse the angular Einstein radius $\theta_{\rm E}$. Here, $\mu_{\text{rel}}$, and $v_{\text{rel}}$ are the lens-source relative proper motion and velocity, while $D_{\rm L},D_{\rm S}$ and $M_{\rm L}$ are the lens distance, source distance and lens mass, respectively. While $t_{\rm E}$ contains information about the lens-mass and lens-source relative distance and velocities, it is completely degenerate. This means that for a single event, $t_{\rm E}$ contains limited information about the lens. However, measurements of $t_{\rm E}$ for large samples of events can tell us about the characteristics of the population of lenses causing them and hence probe the structure of the Galaxy \citep[e.g.,][]{Ki94,EB,Han2003,Kerins2009, Perkins2023}.

While current monitoring surveys of the Galactic bulge like OGLE-IV~\citep{Udalski2015} and KMTNet~\citep{KMTnet2016} have yielded thousands of photometric microlensing events, they are restricted to optical wavelengths. This constrains them to high galactic latitudes ($|b|>2^{\circ}$) around the bulge in windows where extinction from interstellar dust is low. Of course, this does not mean there are no microlensing events in the inner bulge regions. In fact, event rates in the innermost regions of the bulge are expected to be high \citep[e.g.,][]{EB,Specht2020}, they are just not bright enough to be detected at optical wavelengths due to extinction.

\cite{Gould1994} noted that this issue could be overcome by monitoring the inner bulge regions in the Near-Infrared (NIR) wavelengths. NIR photometry is less affected by extinction and provides a means to look through the interstellar dust. The United Kingdom Infrared Telescope (UKIRT) microlensing survey \citep{Shvartzvald2017} was the first to survey the inner regions of the bulge, examining $\approx10$ square degree patch between J$2015$-J$2016$ close to the Galactic plane ($b\approx0$). In addition to supporting space-based microlensing campaigns by both the Kepler and Spitzer space telescopes \citep[e.g.][]{Henderson2016}, the UKIRT microlensing survey reported five highly extinct events missed by optical surveys demonstrating the advantage of observing in the NIR. \edit{Recent analysis of the UKIRT microlensing survey revealed 522 clear events and a lower bound on the event rate increasing towards the bulge \citep{Wen2023}.} Subsequently, efforts to search regions of the inner bulge in the NIR with the Vista Variables in the Via Lactea (VVV) survey \citep{Minniti2010} were undertaken \citep[e.g.,][]{Navarro2017, Navarro2018, NavarroLatitude, NavarroForesaken}. The VVV survey has been monitoring $\approx 560$ squared degrees of sky centred on the Galactic bulge and inner disc in the NIR $K_{s}$-band for nearly a decade. While the VVV may seem ideal for finding microlensing events towards the inner region of the bulge in terms of its wavelength and spatial coverage, it is not a dedicated microlensing survey. Consequently, the adopted sparse and spatially varying cadences makes detecting and characterizing microlensing events difficult. 

Despite the challenges, \citet{Navarro2017} quickly realized the importance of this data set for microlensing. They reported the discovery of $182$ new microlensing events based on observations acquired between J$2010$ and J$2015$ in the three innermost tiles of the survey. This covers an area $1.68^\circ \ge \ell \ge -2.68^\circ$ and $0.65^\circ \ge b\ge -0.46^\circ$.  Later, \citet{NavarroLatitude,NavarroForesaken} extended the microlensing search to cover $14$ tiles, encompassing all of the Galactic plane in VVV. They presented a catalogue of the $630$ microlensing events covering the region within $10.44^\circ \ge \ell \ge -10^\circ$ and $0.65^\circ \ge b\ge -0.46^\circ$. Even so, this is only $14$ of the $348$ tiles in the entire VVV survey. Searching through the whole VVV data set remained an intimidating task. The majority of events found by Narravo et al. searches were dubbed 'forsaken' and not used in subsequent analysis. This was due to the sparse coverage of the events resulting in ambiguous interpretation of the microlensing signals. Specifically, two or more distinct sets of parameters were consistent with the microlensing signal \citep{NavarroForesaken}. These types of events are common in the VVV, so therefore interpreting them is required if the VVV data is to be fully utilized. 

Building on this work, \cite{Husseiniova2021} built a scaleable machine learning classifier to extract microlensing events from the sparse VVV photometry \citep{Smith2018}. They extracted 1959 microlensing events across the entire $560$ deg$^{2}$ VVV survey footprint. \cite{Husseiniova2021} were also able to overcome the problem of the `forsaken' events by using nested sampling \citep{Skilling2006, Higson2019, DYNESTY} to fully characterise the event's multi-modal posterior distributions. \cite{Husseiniova2021} also characterized the spatial dependence on recovery efficiency of their classifier. So, they also were able to compute spatially resolved completeness maps as a function of Einstein crossing time over the entire VVV footprint. \citet{Ka22} subsequently extended this work by incorporating parallax effects into the microlensing models and studied events likely to be caused by dark remnant lenses. In addition to finding photometric microlensing events, there a have also been efforts to predict astrometric microlensing events \citep[e.g.,][]{Zurlo2018, Bramich2018, McGill2020, McGill2023} using NIR data for brown dwarfs lenses \citep{Nielsen2018, Luberto2022} and for sources in the Galactic Bulge \citep{McGill2019}. Most recently, efforts for finding NIR microlensing in the Galactic Bulge have focused on preparations and support for the Roman Space Telescope's Galactic Exoplanet Survey \citep{Bennett2002}, identifying regions of the bulge with the highest event rates in the NIR \citep{Kondo2023, Wen2023} and investigating the microlensing signatures of compact dark matter candidates \citep{Pruett2022, Fardeen2023}.

The most recent and extensive microlensing maps towards the Galactic bulge are based on samples of $>10^{3}$ events from OGLE \citep{Wyrzykowski2015,Wyrzykowski2016,Mroz2019}. Using 3718 events found during $8$ years of the OGLE-III campaign, \cite{Wyrzykowski2015} computed spatially resolved maps of mean Einstein crossing time $\langle t_{\rm E}\rangle$ which were generally in agreement with the Besançon galactic models of the bulge \citep{Kerins2009}. \cite{Wyrzykowski2015} noted the asymmetry in the average timescale $\langle t_{\rm E}\rangle$ around $l=0$, attributed to the structure and viewing angle of the Galactic bar, was slightly more pronounced than expected.  Using a larger sample of 8000 events found during 8 years from the OGLE-IV campaign \citep{Udalski2015}, \cite{Mroz2019} investigated $\langle t_{\rm E}\rangle$, optical depth, and event rate and compared with the second generation Manchester-Besançon Microlensing Simulator \citep[MaB$\mu$lS-2,][]{Specht2020}. In general, \cite{Mroz2019} found agreement with \cite{Wyrzykowski2015} regarding $\langle t_{\rm E}\rangle$ and good agreement with MaB$\mu$lS-2 for predicted $\langle t_{\rm E}\rangle$, optical depth, and event rates.

All current tests of the MaB$\mu$lS-2 have been in the optical wavelengths \citep{SumiPenny2016,Mroz2019} and therefore restricted to low-extinction windows at high galactic latitudes \citep[$1^{\circ}<|b|<6^{\circ}$,][]{Specht2020}. Even though \cite{Navarro2018} and \cite{NavarroLatitude} extracted microlensing events along strips of longitude and latitude in the NIR and covered regions of $|b|<1^{\circ}$, detailed comparisons with Galactic models were not carried out. Regarding the latitudinal strip, \cite{NavarroLatitude} reported an observed decrease in events with increasing latitude and a shorter mean timescale for events found closer to the Galactic plane. For the longitudinal strip, \cite{Navarro2018} noted a smooth increase in the number of events towards the Galactic centre, with a slight asymmetry towards more events at negative latitudes. While the work of \cite{Navarro2018,NavarroLatitude} has demonstrated the power of the VVV for microlensing, work utilizing the full survey is the clear next step. In this paper, we use the work of \cite{Husseiniova2021} and \cite{Ka22} to examine the spatial dependence of the microlensing timescale distribution in the NIR across the Galactic Bulge. 

\begin{figure*}
	\includegraphics[clip, trim=0.0cm 2cm 0cm 2cm,width=2.1\columnwidth]{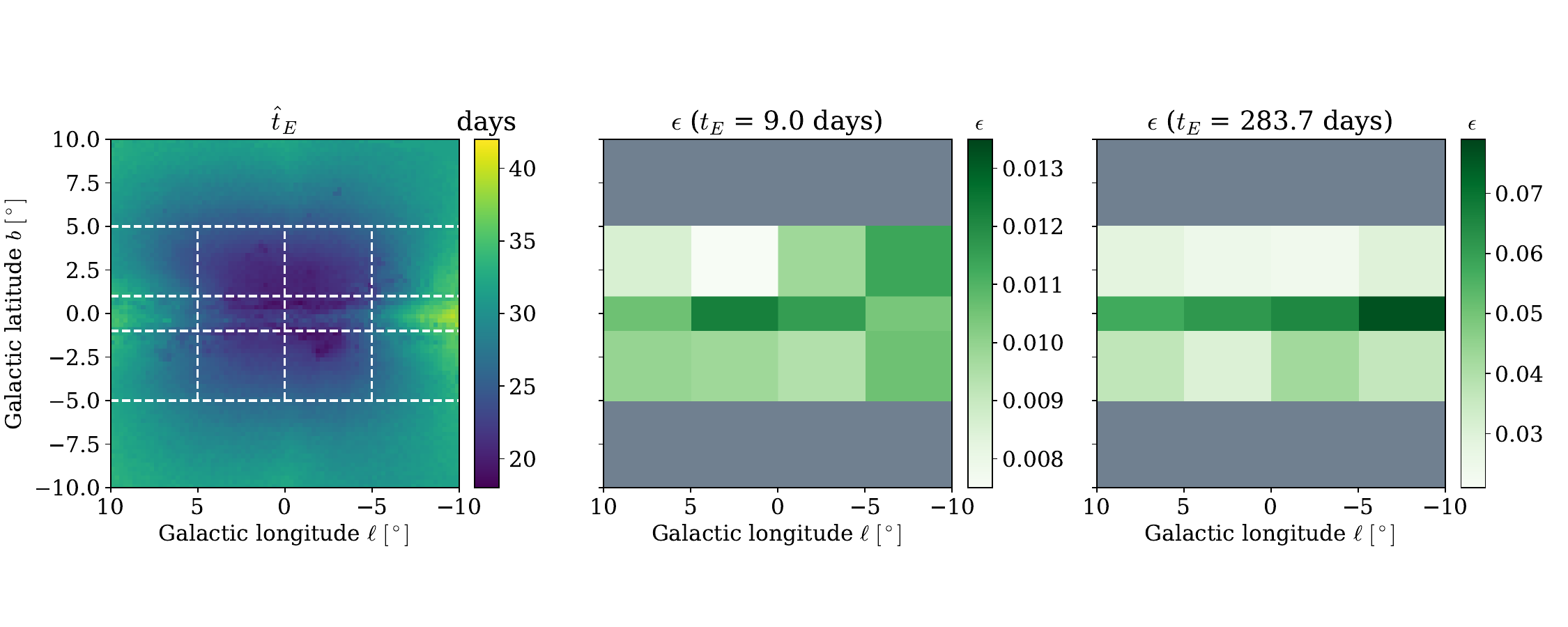}
    \caption{\textit{Left:} Expected \edit{$\hat{t}_{\rm E}$} variation within the adopted spatial bins. White dashed lines denote borders of the bins. Background colours indicate spatial distribution of the expected \edit{mean of fitted lognormal $t_{\rm E}$ distribution, $\hat{t}_{\rm E}$,} in the central regions of the Galaxy in the near-infrared (NIR; K-band). The plot was generated using the second-generation Manchester-Besan\c{c}on Microlensing Simulator \citep[MaB$\mu$lS-2;][]{Specht2020}. \textit{centre and right:} Varying efficiency in recovering microlensing events with a timescale $t_{\rm E} = 9.0$ days (center) and $t_{\rm E} = 283.7$ days (right) under the VVV observing pattern as simulated per each spatial bin. The part filled in grey was not covered by our spatial bins due to insufficient event count.}
    \label{fig:spatial_binning}
\end{figure*}

\section{Methods}

\subsection{Data}

We extract the subsample of 1602 microlensing events limited to the Galactic bulge ($|\ell|<10^\circ,$ $|b|<5^\circ$) from the 1959 total events found by \cite{Husseiniova2021} in version 2 of the VVV infrared astrometric catalog \citep[VIRAC; see][for version 1]{Smith2018}. This catalog provides $K_{s}$-band time series photometry for $<10^{9}$ sources covering $\approx560$ deg$^{2}$ of the Galactic bulge and plane. The VIRAC version 2 catalog was processed with a modified version of DoPHOT \citep{dophot1, dophot2}. The full details of the photometric reduction are described in Smith et al. (in prep). The VVV lightcurves span observations between 2010 to 2019 and have between $\approx50-500$ data points.

\subsection{Einstein timescale posteriors\label{timescale_posteriors}}

We used the posterior distributions obtained by the methods presented in \citet{Ka22}. Specifically, we used nested sampling \citep{Higson2019} implemented via {\tt dynesty} \citep{DYNESTY}, to obtain posterior samples for an annual parallax microlening model parameterized by the Einstein timescale, the normalized lens-source impact parameter, the source blending parameter, the time of lens-source closest approach, $K_s$-band baseline magnitude, and the microlensing parallax vector components in the north and east directions \citep{Gould2004}, or $\{t_{\rm E},u_0,f_{\text{s}},t_{0},m_0, \pi_{\text{EN}}, \pi_{\text{EE}}\}$, respectively. For all of the parameters, flat priors were used which are detailed in Table 1. of \cite{Ka22}. The {\tt dynesty} nested sampling settings were random walk sampling \citep{Skilling2006} with multiple bounding ellipsoids and $1\,000$ initial live points, a stopping criterion in the remaining fractional evidence of $0.01$, and $100$ per cent of weight allocated on computing the posterior distributions. The nested sampling method was chosen due to its ability to sample multi-modal and degenerate posterior distributions, which is important for VVV microlensing events with sparse cadences \citep{Husseiniova2021, Ka22}.

\subsection{Hierarchical timescale inference \label{timescale_inference}}

To infer the $t_{E}$ distribution for a sample of events, we follow \cite{Golovich2020} and fit flexible step function to the distribution, $f_\alpha$, within a hierarchical model of the form,
\begin{equation}
    f_{\boldsymbol{\alpha}}(\log_{10}t_{E})=\sum_{m=1}^{M}\alpha_{m}s\left(t_{E};\frac{m-1}{M},\frac{m}{M}\right),
\end{equation}
where $\boldsymbol{\alpha}$ is a vector of length $M$, with the elements $\alpha_{m}$ and,
\begin{equation}
    s(x, L, H) = 
    \begin{cases}
    0 & \text{for } x < L, \\
    (H - L)^{-1} & \text{for } L\leq x\leq H,\\
    0 & \text{for } x > H
    \end{cases}
\end{equation}
where we divide the allowed range of event timescales into $M=30$ log-space bins between $\tE=1$ and $\tE=10^3$ days, and specify that,
\begin{equation}
    \sum_{m=1}^{M}\alpha_{m}=1,\label{eq:normalized}
\end{equation}
to ensure that $f_\alpha$ is a normalized probability distribution. Under this model we can compute the likelihood efficiently and approximately over a sample of microlensing events for a value of $\boldsymbol{\alpha}$ by using importance sampling and re-weighting the posterior samples for each event obtained by the methods described in Section \ref{timescale_posteriors}. For a sample of $N_{s}$ events where $t_{E_{n,k}}$ is the kth posterior sample of the nth event, the likelihood is \citep{Hogg2010}, 

\begin{equation}
\label{likelihood_alpha_final}
\mathcal{L}({\{d_{i}\} |\boldsymbol{\alpha})\approx\prod_{n=1}^{N_{s}} \frac{1}{K} \sum_{k=1}^{K} \frac{f_\alpha(\tEnk)}{p_0(\tEnk)}\textnormal{.}
}
\end{equation}

Here, $\{d_{i}\}$ is the set of $N_{s}$ microlensing lightcurves and $p_{0}$ is the prior density from Table 1. in \cite{Ka22}. We place a Dirichlet prior on $\boldsymbol{\alpha}$ to enforce the normalization condition in Eq. (\ref{eq:normalized}),
\begin{align}
p(\boldsymbol{\alpha}) \sim \textnormal{Dir} (\boldsymbol{\alpha}|\boldsymbol{a}) \textnormal{, where: } a = (\boldsymbol{w} - \min(\boldsymbol{w}))b + c
\end{align}
Here, following \cite{Golovich2020}, we set the concentration hyper-parameters, $\boldsymbol{a}$, using the weights of the maximum a posteriori (MAP) $\tE$ histogram ($\boldsymbol{w}$) and we fix $b=20$, $c=0.2$. We then sample the posterior distribution of $\boldsymbol{\alpha}$ using the No U-Turn Sampler (NUTS) algorithm \citep{NUTS} implemented in {\tt PyMC3} \citep{PyMC3} using $1600$ burn-in steps and then a total of $3200$ steps for inference.

\begin{table}
\centering
\begin{tabular}{c|c|c|c} 
\textbf{bin 1} & \textbf{bin 2} & \textbf{bin 3} & \textbf{bin 4} \\
76 events & 139 events & 134 events & 43 events \\[5pt]
\hline
\textbf{bin 5} & \textbf{bin 6} & \textbf{bin 7} & \textbf{bin 8} \\
122 events & 268 events & 247 events & 113 events \\[5pt]
\hline
\textbf{bin 9} & \textbf{bin 10} & \textbf{bin 11} & \textbf{bin 12} \\
75 events & 190 events & 143 events & 52 events \\

\end{tabular}
\caption{\edit{Breakdown of the 1602 VVV microlensing events used in this analysis: event count per spatial bin and bin numbering for future reference.} The 4x3 grid matches that of Fig~\ref{fig:spatial_binning}.}
\label{table:bin_division}
\end{table}

\edit{\subsection{Choice of metric}
\label{sec:metric}
Throughout this work we operate in the $\log t_{\rm E}$ scale. Following \citet{Wyrzykowski2015}, we fit the event timescale distribution with a lognormal model; we use the mean of this lognormal distribution (which we denote as $\hat{t}_{\rm E}$) as a metric representing a typical event timescale for a given bin (e.g. Fig~\ref{fig:spatial_binning}). This metric is more robust than the mean timescale $\langle t_{\rm E} \rangle$, as the $\log t_{\rm E}$ distribution is roughly symmetric; the peak of this distribution should be minimally sensitive to outliers. This is especially important for the part of this work concerning VVV data, where outer wings of the $t_{\rm E}$ distribution are dominated by noise. We describe the fitting procedure in more detail in Section \ref{sec:lognormal}.}

\subsection{The Manchester-Besan\c{c}on Microlensing Simulator (MaB{\textmu}lS-2) \label{mabuls}}

To compare our microlensing timescale distributions to models of the Galaxy we use the second version of the Manchester-Besan\c{c}on Microlensing Simulator \citep[MaB{\textmu}lS-2;][]{Specht2020}, which is the current state-of-the-art tool for creating maps of microlensing event population observables -- optical depth, event rate and timescale distributions -- along different lines of sight towards the Galactic Bulge. MaB{\textmu}lS-2 is publically available\footnote{\url{www.mabuls.net}} and covers a $20^\circ$ x $20^\circ$ region around the Galactic Centre and provides simulations for several optical and NIR bands. Based on the Besan\c{c}on stellar population synthesis model \citep[][]{Robin2003, Robin2012}\footnote{\url{https://model.obs-besancon.fr/}}, MaB{\textmu}lS-2 simulations include thin and thick disk, bar and halo components which all have individually defined density profiles, initial mass functions, star formation rates, kinematics and stellar ages \citep[see Section 2 of][]{Specht2020}. Additionally, interstellar dust is accounted for in the simulations using a 3D Galactic extinction model \citep{Marshall2006}.

MaB{\textmu}lS and the underlying Besan\c{c}on Galactic model have been iteratively improved when reconciled with data. For example, analysis of microlensing events from MOA-II \citep{Sumi2013} led to additional populations of faint, low mass M dwarfs and brown dwarfs being added to the simulations \citep{Mabuls1,SumiPenny2016}. Most recently, motivated by upcoming large scale time-domain surveys such as the \textit{Roman Space Telescope} \citep{Spergel2015} and the \textit{Vera C. Rubin Observatory} \citep{Ivezic2019}, MaB{\textmu}lS-2 includes the simulation of higher-order microlensing effects such as finite sources, unresolved stellar backgrounds and is in good agreement with the largest samples of $\sim10^{4}$ optical microlensing events from OGLE-IV \citep{Udalski2015,Specht2020} and stellar kinematics observed by the \textit{Hubble Space Telescope} towards the bulge.

\subsection{Spatial mapping \label{timescale_spatial_mapping}}

We cover the region of interest around the Galactic Centre ($|\ell|<10^\circ,$ $|b|<5^\circ$)  with 12 rectangular bins. Those include 4 narrow bins around the Galactic plane ($-1^\circ<b<1^\circ$, uniform $5^\circ$ steps in longitude between $-10^\circ$ and $10^\circ$), and 8 wider bins on either side ($-5^\circ<b<-1^\circ$ and $1^\circ<b<5^\circ$, uniform $5^\circ$ steps in longitude between $-10^\circ$ and $10^\circ$). This choice of binning is motivated by the spatial variation of event timescales expected from Galactic models, as illustrated in Fig~\ref{fig:spatial_binning} and Table~\ref{table:bin_division}, respectively. In selecting our bins, we aim to both capture the most dramatic features of the event timescale map predicted with MaB{\textmu}lS-2 (e.g. ridge at $b\approx 0 ^\circ$, sharp rise in timescales around $\ell\approx\pm 7.5 ^\circ$), and test whether we can detect any asymmetry in the Bulge between positive and negative longitudes. The bin selection was also optimised to contain a sufficient number of events in every subsample.

\subsection{Efficiency correction \label{timescale_efficiency_mapping}}

The efficiency of recovering events with various timescales is imposed by the VVV observing pattern and cadence and is highly heterogeneous across the survey area \citep[see Figure 8 in][]{Husseiniova2021}. After obtaining hierarchical $\tE$ distributions for each of the bins, we need to correct them for this efficiency; neglecting this step would introduce additional bias being solely the consequence of the particular VVV observing program used.

To generate efficiency correction maps, we adopt an approach similar to the one described in Section 4.3 of \citet{Husseiniova2021}. The area around the Galactic Bulge is divided into rectangular bins, and microlensing events are simulated in each bin following a point source -- point lens model assuming linear relative motion (no parallax effect). A source 'seed' -- effectively, the position of the event, its observation times and source star baseline magnitude -- is selected randomly using real light curves from within the area. The timescale $t_{\rm E}$, for which we are determining the recovery efficiency, is fixed. The remaining model parameters $u_0$, $f_s$, $t_0$ are selected randomly from distributions described in Table~\ref{table:efficiency_map_param_distributions}. The simulated light curves are then passed through the automatic classifier developed in \citet{Husseiniova2021}, using a threshold probability of 0.825 required for event recovery, until the recovery efficiency reaches the signal-to-noise ratio of 10.

\begin{table}
\centering
\begin{tabular}{cll} 
\hline
parameter & PDF & unit \\
 \hline
 $u_0$ & uniform($10^{-3}$, 1.5) & $\theta_E$ \\
 $f_{\text{s}}$ & uniform(0, 1) & - \\
 $t_{0}$ & uniform(2019, 2021) & Julian years \\

\end{tabular}
\caption{Model parameter probability distributions used for generating random events in the process of simulating microlensing event recovery efficiency in the VVV.}
\label{table:efficiency_map_param_distributions}
\end{table}
\begin{figure}
\includegraphics[width=1.1\columnwidth]{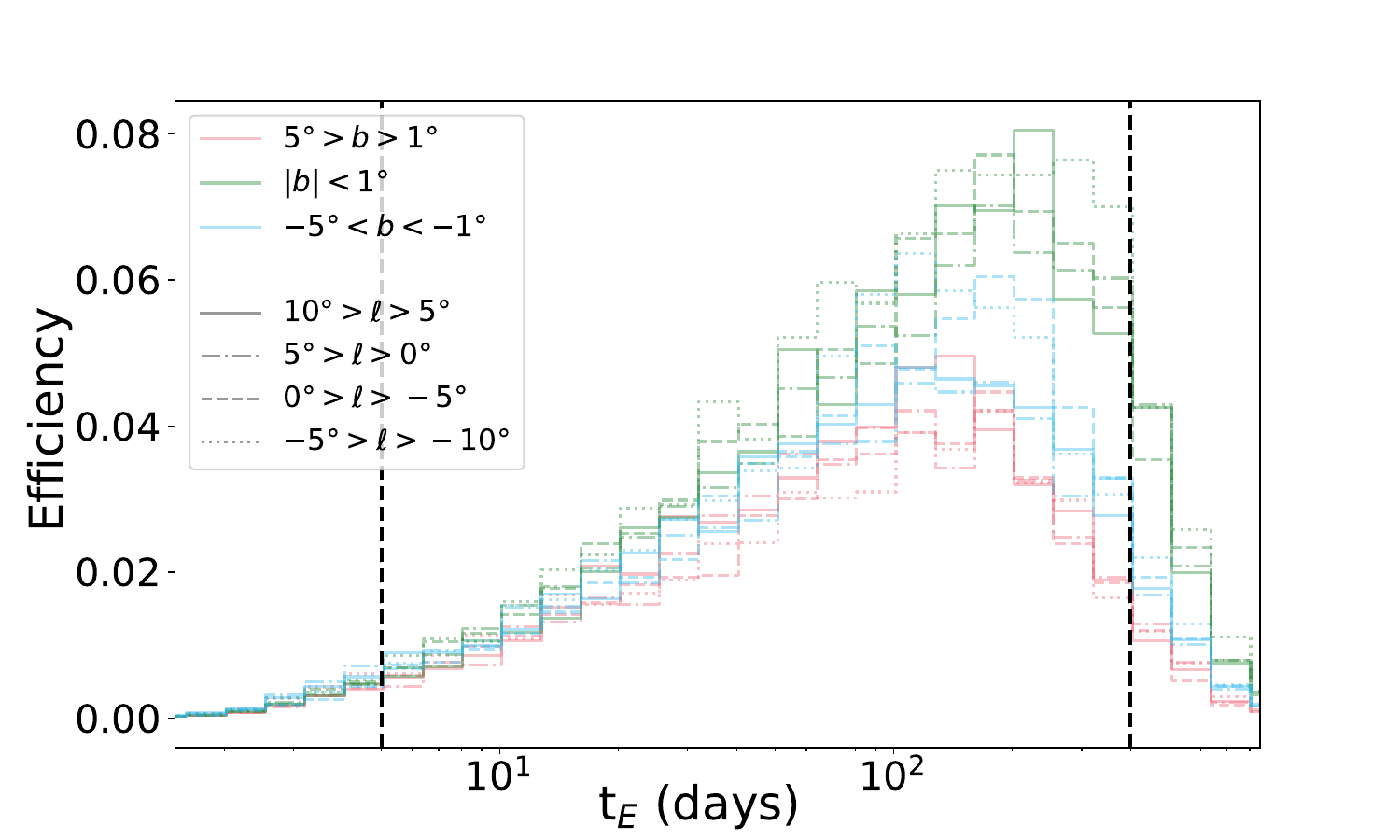}
\caption{Varying efficiency (estimated with error $\sigma_\epsilon \lesssim 0.1 \epsilon$) in recovering microlensing events as a function of timescale $t_{\rm E}$ under the VVV observing pattern, as simulated per each spatial bin. Colours represent latitude intervals \edit{and linestyles represent longitude intervals}; 
the dashed vertical lines denote the boundaries of the mask described in Subsection \ref{sec:lognormal}.}
\label{fig:efficiency_curves}
\end{figure}

The main difference is that our efficiency map generation adopts a higher resolution in $\tE$, but a lower spatial resolution. Adopting a coarse grid ($2^\circ$ x $5^\circ$ to $4^\circ$ x $5^\circ$ regions) matching the spatial bins makes the process significantly less computationally expensive; within each of the bins, the simulation was run for each of the 30 midpoints of $\tE$ bins used in the hierarchical $\tE$ distribution inference. This allows for straightforward correction of the height of each $\tE$ bin by its corresponding recovery efficiency in the spatial bin.

We find that the efficiency curves between our spatial bins are significantly different. This is mostly the effect of the specific VVV observing pattern \citep[see Figure 7 in][]{Husseiniova2021}. The differences in efficiency curves mostly reflect two features of this pattern: longer survey coverage (higher standard deviation in observation date) at low latitudes, and a small window around $\ell = 5^\circ$, $b = 2,5^\circ$ having a distinctly higher cadence. We show the spatial efficiency variations for two sample timescales in the middle and right panels of Fig~\ref{fig:spatial_binning}.  A small, though notable, asymmetry in latitude is also reflected in the efficiency curves, especially in the wing corresponding to longer events, as illustrated in Fig~\ref{fig:efficiency_curves}.

\begin{figure*}
	\includegraphics[clip, trim=2cm 6.5cm 2cm 0cm, width=1.7\columnwidth]{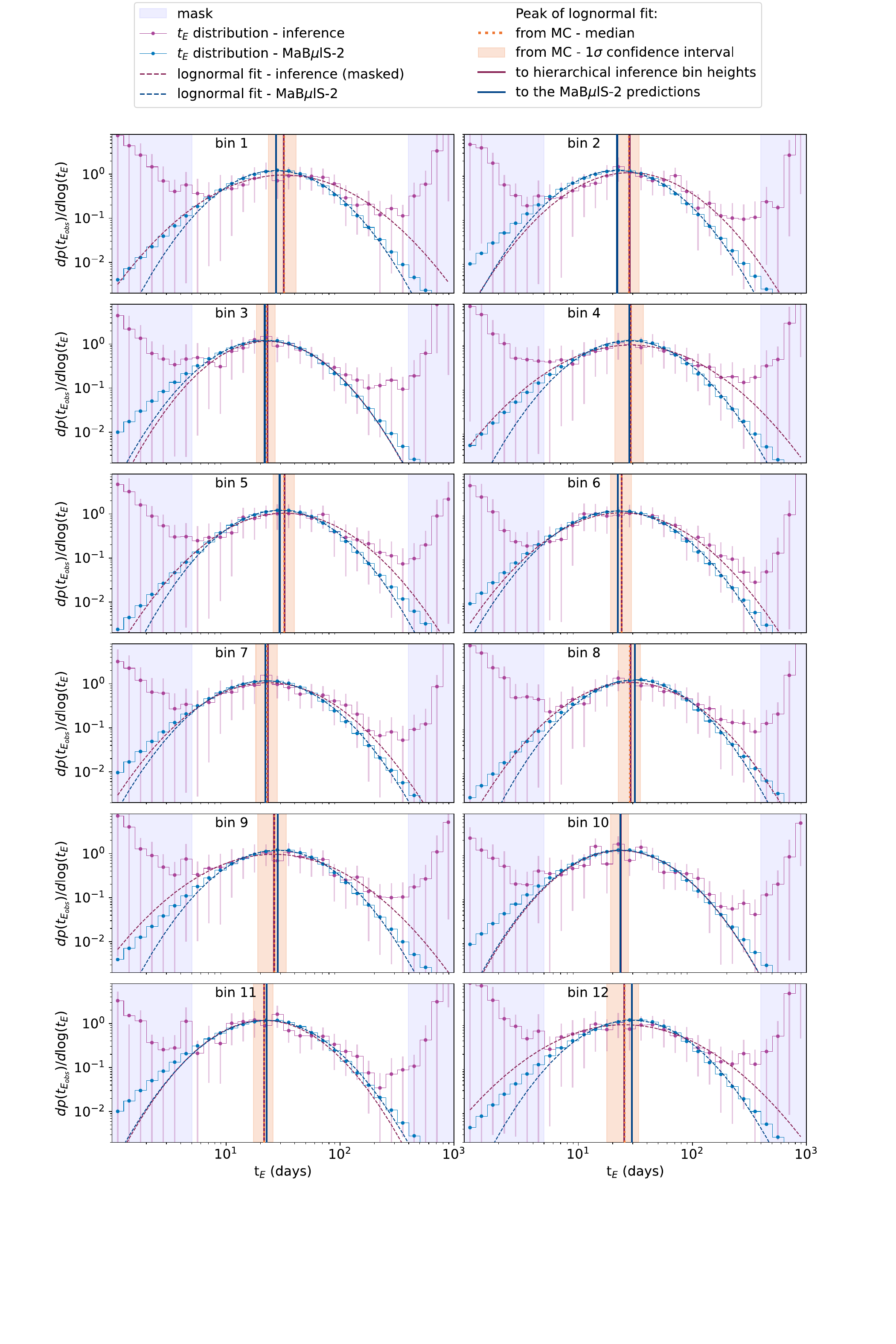}
  \caption{Event timescale distributions for all 12 spatial bins. Purple, solid line/filled circles: efficiency-corrected $\tE$ bin heights from VVV events obtained using hierarchical inference. Errorbars indicate 68\% confidence intervals. Blue, solid line/filled circles: $\tE$ bin heights predicted using the MaB{\textmu}lS-2 simulation. Dark purple: lognormal distribution fit to datapoints obtained using hierarchical inference. Only the datapoints outside of the mask (light blue, filled) are included in the fit. The dashed line is the fitted curve and the vertical line is the mean ($\hat{t}_{\rm E}$) of the distribution. Dark blue: as dark magenta, but fitted to MaB{\textmu}lS-2 datapoints. Orange: median (dotted line) and  68\% confidence intervals (light orange band) for the means ($\hat{t}_{E,i}$) of the lognormal distributions fitted to each histogram $y_i(\log t_{\rm E})$ from 10000 MC samples.}
  \label{fig:timescales_combined}
\end{figure*}
\begin{figure*}
	\includegraphics[clip, trim=1cm 0cm 0cm 0cm, width=2\columnwidth]{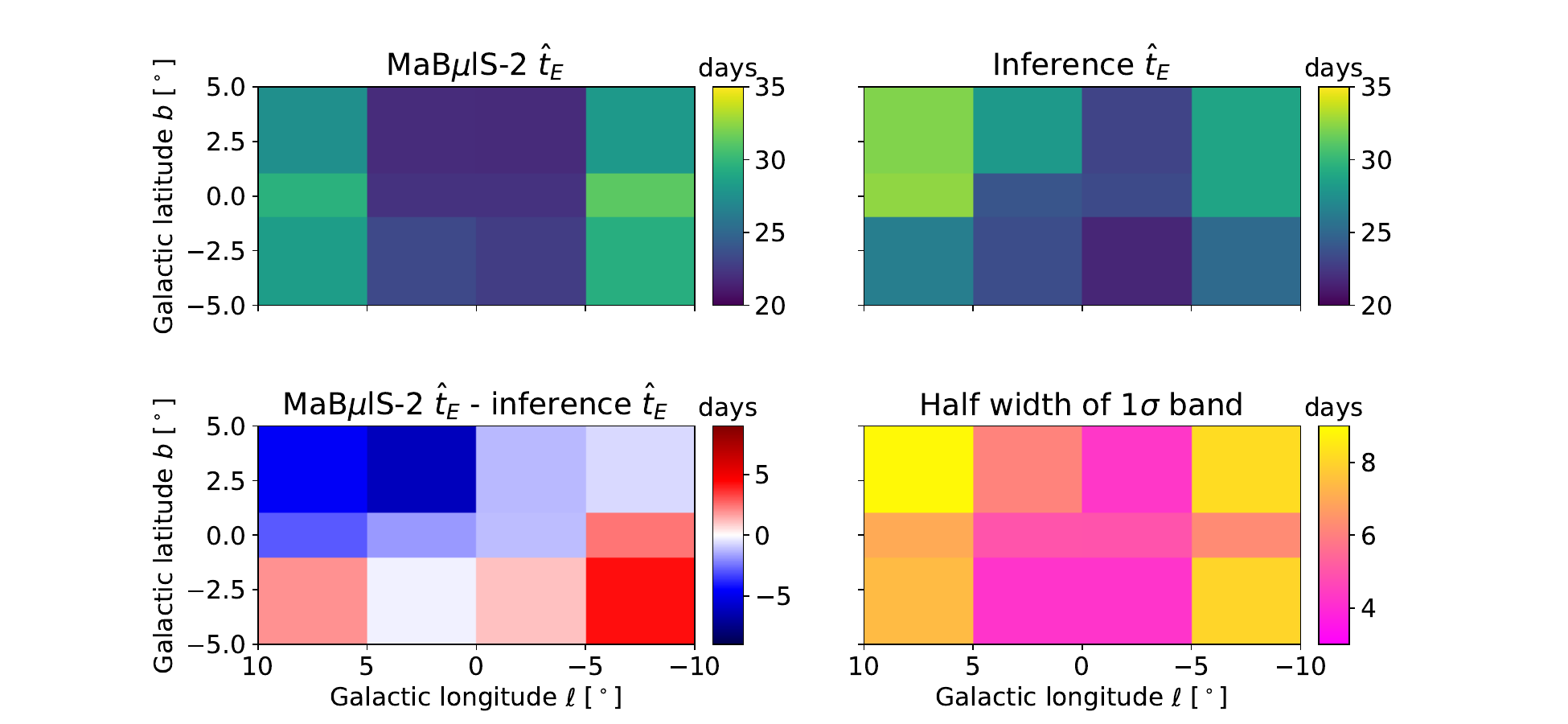}
  \caption{Timescales $\hat{t}_{\rm E}$ corresponding to the means of the fitted lognormal timescale distribution for each bin. \textit{Top left:} fit to MaB{\textmu}lS-2 simulation predictions. \textit{Top right:} fit to hierarchical inference results. \textit{Bottom left:} Difference between predicted and inferred $\hat{t}_{\rm E}$. \textit{Bottom right:} Half width of the 1$\sigma$ confidence interval for determining $\hat{t}_{\rm E}$ from hierarchical inference results (see Fig~\ref{fig:timescales_combined}). The dimensions match those of Fig~\ref{fig:spatial_binning}; only regions covered by the bins included.}
  \label{fig:timescales_colormap}
\end{figure*}
\begin{figure*}
	\includegraphics[clip, trim=2cm 0.5cm 3cm 4.8cm,width=1.9\columnwidth]{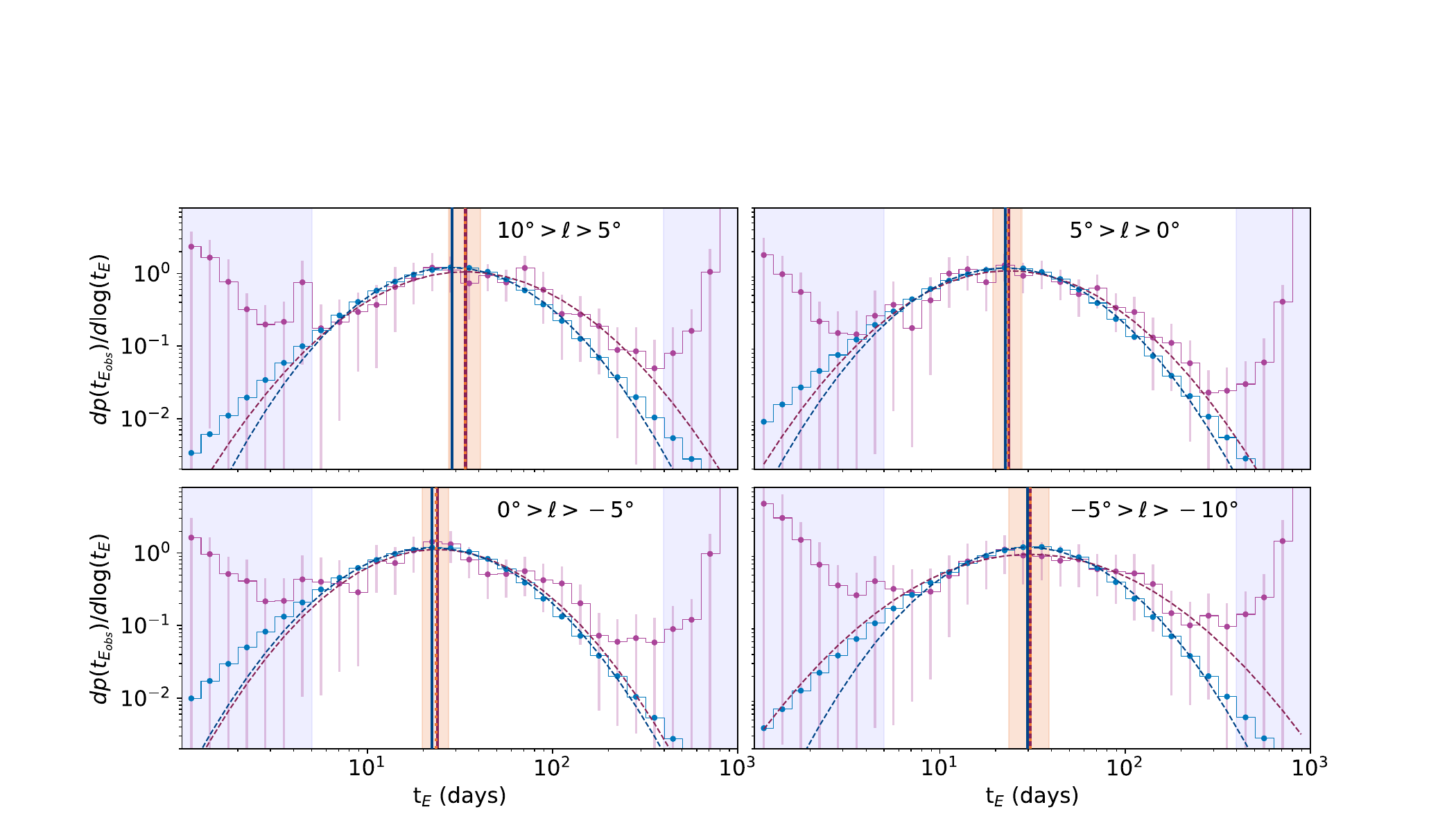}
    \caption{Event timescale distributions. As Fig~\ref{fig:timescales_combined}, plotted per each longitude spatial bin; legend as in Fig~\ref{fig:timescales_combined}.}
    \label{fig:timescales_lon}
\end{figure*}
\begin{figure}
	\includegraphics[clip, trim=0.5cm 0.7cm 0cm 0cm,width=1\columnwidth]{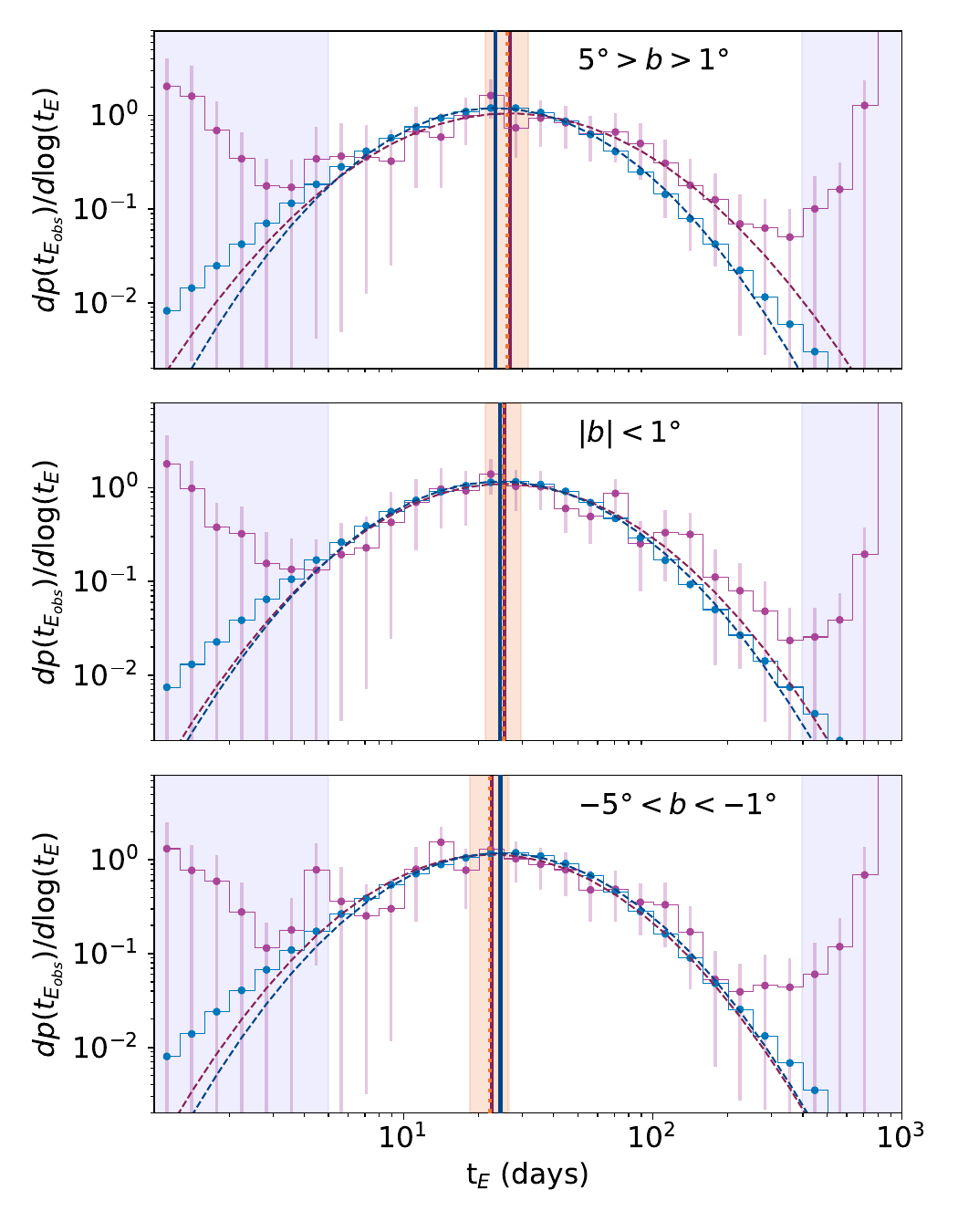}
    \caption{Event timescale distributions. As Fig~\ref{fig:timescales_combined}, plotted per each latitude spatial bin; legend as in Fig~\ref{fig:timescales_combined}.}
    \label{fig:timescales_lat}
\end{figure}
\begin{figure*}
	\includegraphics[clip, trim=3cm 0cm 2.5cm 0cm, width=2.1\columnwidth]{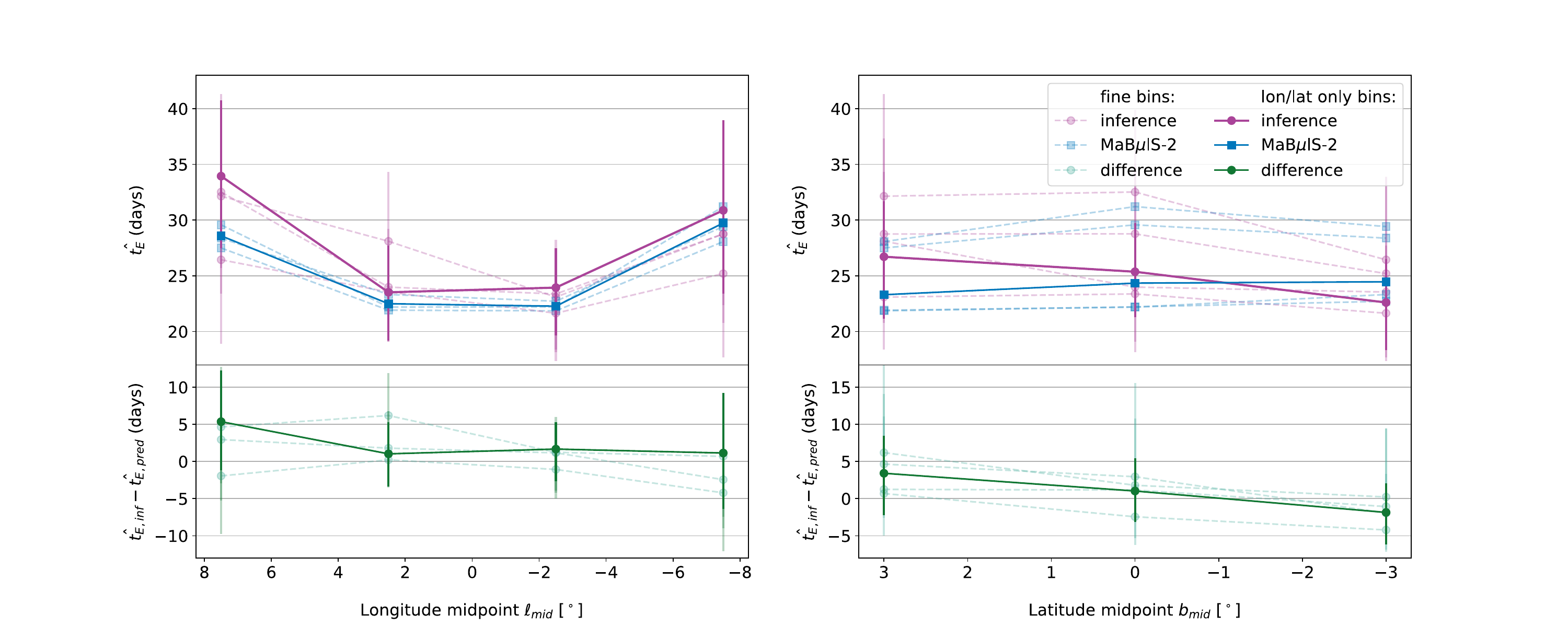}
    \caption{Inferred and predicted (MaB{\textmu}lS-2) peak timescale distribution values $\hat{t}_{\rm E}$ as a function of longitude (left) and latitude (right). Purple circles: hierarchical inference results. Blue squares: MaB{\textmu}lS-2 predictions. Green circles: difference between inference and MaB{\textmu}lS-2 values for respective bins. Dark, solid: for the longitude-only binning. Light, dashed: for each of the latitude intervals separately. Errorbars indicate 68\% confidence intervals.}
    \label{fig:lon_lat_diff}
\end{figure*}

\subsection{\edit{Limitations of the efficiency correction}}
\label{sec:limitations}
\edit{Our efficiency correction is subject to several limitations:

\begin{enumerate}
    \item At the stage of creating mock events for the efficiency correction, we assume no parallax signal ($\pi_E = 0$); whereas for the hierarchical inference, to obtain maximally precise timescale measurements, we use samples from modelling that includes parallax effect. \citet{Ka22} note that 176 out of 1959 modelled events (9.0\%) had strong evidence for parallax signal (at the level of Bayes factor $K>100$). The majority of the events with strong parallax signal have very high inferred event timescales ($t_E > 400$ days), but much shorter inferred timescales from the no-parallax modelling ($t_E < 400$ days). For such an event, its detection efficiency would be underestimated (to be rightward of the black dashed line in Fig. \ref{fig:efficiency_curves}) and hence, the corresponding bins disproportionally upweighted. As it is not feasible to disentangle contributions of single events to the hierarchical inference, we also cannot amend the efficiency correction on a per-event basis. We consider this the main shortcoming of the efficiency correction model. \\
    After analysing this subset of 176 events in more detail, we typically find very high uncertainties in timescale and parallax fits, as well as lightcurves with sparse cadence and only a single slope of the event covered. We conclude a significant fraction of those objects could be contaminants, mainly long-period variables whose other peaks are missed due to sparse cadence; however their nature (microlensing or intrinsic variable) cannot be unambiguously determined with the available data. This also motivates more strict masking of the long-timescale wing of the $t_{\rm E}$ distribution (see Subsection \ref{sec:lognormal}). The contamination problem should be, to a large degree, alleviated with next-generation high-cadence NIR observations. \\

    \item At the stage of creating mock events, we assume a uniform distribution of the blending parameter $f_{\rm s}$ \citep[following][]{Husseiniova2021}.
    The $f_{\rm s}$ distribution in the $K_{\rm s}$ band is not known as of today due to insufficient data; we expect this obstacle to be entirely eliminated in the near future with data from JWST and Roman Space Telescope. We note the possibility for post-correction of our hierarchical inference results once high-resolution images in the NIR are available for determining the $f_{\rm s}$ distribution.\\

    \item The final catalogue of 1959 events from \citet{Husseiniova2021} has been filtered with human inspection, which we do not account for in the efficiency correction model. However, \citet{Husseiniova2021} do not note a correlation of completeness of the visual inspection with event timescale (they note a correlation with baseline magnitude and impact parameter). We thus conclude this should not introduce a significant bias.
\end{enumerate}}

\section{Results} \label{results_section}
\subsection{Obtaining the $t_{\rm E}$ distributions \label{te_first}}

For all 12 spatial bins, we obtain the $t_{\rm E}$ via hierarchical inference (Subsection \ref{timescale_inference}), simulate the detection efficiency $\epsilon$ (Subsection \ref{timescale_efficiency_mapping}) and correct for it by multiplying every $t_{\rm E}$ distribution bin height by $1/\epsilon$, and then compare the results to the predictions from MaB{\textmu}lS-2 (Subsection \ref{mabuls}). Those predictions were obtained using a custom run of the simulator modified to output full $t_{\rm E}$ distributions -- all input and settings are the same as in \citet{Specht2020} and the online simulator tool\footnote{\url{www.mabuls.net}}. We interpolate $t_{\rm E}$ distributions from the simulation in each MaB{\textmu}lS-2 spatial bin ($0.25^\circ$ x $0.25^\circ$) to match our $t_{\rm E}$ binning (30 logspace bins between $t_{\rm E} = 1$ and $t_{\rm E} = 10^3$ days), and merge those bins (weighted by event rate) to match our spatial binning. The efficiency-corrected timescale distributions from the hierarchical inference, along with their respective MaB{\textmu}lS-2 predictions, are shown in Fig~\ref{fig:timescales_combined} for all the fine bins.

For clarity, we do not include the hierarchical inference results before the efficiency correction. To give an idea of the corrections for bins with different typical efficiencies, we present the non-corrected distributions from bins 1, 5 and 9 (leftmost column) in Fig~\ref{fig:timescale_correction} and discuss them in Appendix \ref{sec:non-corr}.

The errorbars in Fig~\ref{fig:timescale_correction} and Fig~\ref{fig:timescales_combined} indicate the 68\% confidence intervals on the $\tE$ bin heights. The non-corrected errorbars are taken directly from the samples of $\tE$ bin heights from the hierarchical inference code, while the efficiency-corrected distribution errorbars also include the error on the recovery efficiency value. The extents of those errorbars were calculated with a Monte Carlo simulation, drawing 100,000 efficiency-corrected bin height values per each $\tE$ bin with both a random choice of the bin height from the samples and a random choice of the efficiency correction from a normal distribution $N(\epsilon, \sigma_{\epsilon})$. ($\sigma_\epsilon$ is included in the output of the detection efficiency simulator and is determined by the target signal-to-noise ratio criterion: $\sigma_\epsilon \lesssim 0.1 \epsilon$.)

\subsection{Lognormal distribution fitting}
\label{sec:lognormal}

\edit{As discussed in Section \ref{sec:metric}, we fit a lognormal distribution of $t_{\rm E}$ and use the mean of this distribution, $\hat{t}_{\rm E}$, as a metric.

At the preliminary stages of data analysis, we find very high noise levels in $\hat{t}_{\rm E}$ determination (\newedit{see} Appendix \ref{sec:non-corr}) and we note that the outer wings of the distribution contain unrealistically large best-estimate bin heights, albeit with very large errorbars. This shape of the $t_{\rm E}$ distribution is in disagreement with \citet{MaoPaczynski1996}, \newedit{who argue that the wings of the event timescale distribution should be power laws.} \newedit{\citet{MaoPaczynski1996} explain that long timescale events are produced by lenses with small transverse velocity (if the transverse component is significantly lower than the velocity dispersion for the three-dimensional Gaussian velocity distribution, they derive the number of events to be N $\sim t^{-3}$). On the other slope, the short events are produced by the lenses close to the source or the observer; by expressing the event timescale as a function of lens distance, they arrive at N $\sim t^{3}$.}

We find several possible causes for this effect. Firstly, the outer wings of the distribution from the hierarchical inference typically flatten out smoothly, yielding larger bin height values than histograms from raw samples or simulation predictions \citep[compare Figures 11, 12 in][]{Golovich2020}. Secondly, the detection efficiency simulated for the VVV observing strategy is very low for the lowest and highest $t_{\rm E}$ values, resulting in significant upweighing during efficiency correction. This means any overestimation in the outer wings of $t_{\rm E}$ distribution will be highly amplified.

This effect is significantly stronger, and the slope steeper, for the long $t_{\rm E}$ end, despite the efficiency detection being higher there. \newedit{With visual inspection,} we find that a large fraction of very long events ($t_{\rm E}$ > 400 days) appear to be ambiguous and could be consistent with both microlensing and intrinsic variables (see also Section \ref{sec:limitations}).

\newedit{In order to estimate the degree of ambiguity with minimal bias, we extract quality grades given by the authors in \citet{Husseiniova2021} during visual inspection and define average score per author < 1 to be an indication of ambiguity.
Using the median $t_{\rm E}$ from posterior samples, we find 34\% of very long ($t_{\rm E}$ > 400 days) events and 33\% of very short ($t_{\rm E}$ < 5 days) events from our dataset received such grades, as compared to 19\% of events outside those ranges.}

We conclude the outer wings of our analysis are disturbed by noise, coming primarily from non-microlensing contaminants, which is highly amplified at the later stages. (Second-order effects could also include degeneracies leading to inaccurate timescale determination, e.g. \citet{Gould2004}, or inaccurate efficiency correction for real microlensing events with very large differences between non-parallax and parallax $t_{\rm E}$ estimates; however, we estimate their impact to be much smaller.) To counter this disturbance, we apply a mask that excludes bins of $t_{\rm E}$ < 5 days and $t_{\rm E}$ > 400 days (7 bins at the low $t_{\rm E}$ edge and 4 bins at the high $t_{\rm E}$ edge). We overplot those boundaries on the efficiency curves (Fig~\ref{fig:efficiency_curves}). The mask boundaries were chosen to be maximally wide so that $y(\log t_{\rm E})$ (where $y = dp(t_{E_{\rm obs}}) / d\log(t_{E})$) remains concave within the extent for most bins. We note our choice of metric $\hat{t}_{\rm E}$ minimises the impact of the arbitrary choice of a mask on the results, as the peak should be uniquely determined and insensitive to the outer wings (unlike e.g. $\langle t_{\rm E} \rangle$ or $10^{\langle \log  t_{\rm E} \rangle}$). We also emphasize that the need for applying the mask is imposed by data quality; we expect it to be entirely alleviated in upcoming high-cadence surveys with higher microlensing event yield and sample purity.

We normalise the efficiency-corrected distribution to 1 using only the non-masked values before over-plotting it on the MaB{\textmu}lS-2 predictions in Fig~\ref{fig:timescales_combined} (in practice, assigning bin heights = 0 to the masked $t_{\rm E}$ bins). We also apply the same correction to the MaB{\textmu}lS-2 bin heights. After applying this normalisation, almost all the non-masked bin heights agree within 1$\sigma$ with the MaB{\textmu}lS-2 predictions, though there are a few outliers.

We fit a lognormal distribution using {\tt scipy.optimize.curve\_fit} \citep{scipy} to the unmasked extents of the $t_{\rm E}$ distributions from both the hierarchical inference results from VVV data and the MaB{\textmu}lS-2 predictions. We present the fits alongside individual distributions in Fig~\ref{fig:timescales_combined}, and we compare the $\hat{t}_{\rm E}$ values per bin on a colour map in Fig~\ref{fig:timescales_colormap}. The systematic differences between predicted and inferred $\hat{t}_{\rm E}$ seen in those results are discussed in detail in Subsection \ref{trends}.}

\subsection{Confidence intervals}

To evaluate the precision of our $\hat{t}_{\rm E}$ estimation from data, we run a Monte Carlo simulation. For each spatial bin we apply the following procedure 10000 times. We draw a set of efficiency-corrected $t_{\rm E}$ bin heights randomly out of the Monte Carlo samples obtained in Subsection \ref{te_first}, generating a single realisation (with index $i$) of a $t_{\rm E}$ histogram: $y_i(\log t_{\rm E})$. For this realisation, we fit a lognormal distribution and save its mean $\log \hat{t}_{{\rm E},i}$.

In Fig~\ref{fig:timescales_combined}, we plot the median values of $\hat{t}_{{\rm E},i}$ and their 68\% confidence intervals, along with 1000 randomly chosen MC samples $y_i$, for each bin. We also plot the $\hat{t}_{\rm E}$ values for both our inference and MaB{\textmu}lS-2 with vertical lines. We conclude that our inference results are consistent with MaB{\textmu}lS-2 predictions within 1$\sigma$, except for bin \edit{2} (21.\edit{9} days) which is just outside the 1 sigma band (22.1 - 34.2 days).  We plot the half width of the 1$\sigma$ confidence interval as part of Fig~\ref{fig:timescales_colormap} (bottom right). This width is very highly correlated with event count per bin (see Table \ref{table:bin_division}).

\subsection{Trends in longitude and latitude}
\label{trends}
In addition to the fine binning, we then also repeat all steps outlined in Subsection \ref{te_first} for the merged longitude-only binning ($5^\circ > b > -5^\circ$ and, respectively: $10^\circ >\ell> 5^\circ$, $5^\circ >\ell> 0^\circ$, $0^\circ >\ell> -5^\circ$, $-5^\circ >\ell> -10^\circ$) and latitude-only binning ($10^\circ >\ell> -10^\circ$ and, respectively: $5^\circ > b > 1^\circ$, $|b| < 1^\circ$, $-5^\circ < b < -1^\circ$) in order to maximise the signal-to-noise ratio and specifically search for asymmetry. The timescale distributions, analogous to Fig~\ref{fig:timescales_combined}, are shown in Figs~\ref{fig:timescales_lon} and \ref{fig:timescales_lat} for the longitude and latitude bins, respectively.

\edit{We note that the errorbars on $\hat{t}_{\rm E}$ shrink slightly when adapting the coarser binning, with the mean difference between the upper and lower bounds of the 68-percentile extent of all MC samples equal to 12.4 days for the fine binning, 11.1 days for the longitude-only binning and 8.9 days for the latitude-only binning.} The differences between bins are still not statistically significant with the coarser binning.

To analyse the trends reflected in our estimates and biases influencing them, we plot the estimated peaks of the timescale distribution $\hat{t}_{\rm E}$ as a function of longitude only and latitude only in Fig~\ref{fig:lon_lat_diff}.

\edit{As compared to MaB{\textmu}lS-2 predictions}, we reproduce the trend of lower timescales in central longitudes ($|\ell| < 5^\circ$) and higher timescales in non-central longitudes ($|\ell| > 5^\circ$), but we do not reproduce the trend of timescales increasing with decreasing latitude. Our estimates are slightly biased towards longer timescales -- averaging the differences weighted by $1/\sigma^2$, we obtain a bias of \edit{+0.7} days for the fine binning (\edit{+1.9} days for longitude binning and \edit{+0.4} days for latitude binning, respectively). The differences are higher for higher latitudes and higher longitudes.

\edit{With inverse variance weighing over the fine binning, we obtain an average peak of the lognormal timescale distribution at 23.6 $\pm$ 1.9 days for the central bins and 29.0 $\pm$ 3.0 days for the non-central bins. This is in good agreement with the MaB{\textmu}lS-2 predictions of 22.4 days and 29.0 days (averaged with equal weights per bin), respectively.}

\section{Discussion and Conclusions}

We analyzed a sample \edit{of} 1602 microlensing events found in the VVV data and obtained spatially-resolved timescale distributions across the Galactic bulge, using hierarchical inference with all nested-sampling posterior samples from modelling the events as input. We then compared those timescale distributions with MaB{\textmu}lS-2 predictions. Therefore, we provide a new test of the MaB{\textmu}lS-2 simulation (and, indirectly, the Besançon model),  independent of previous studies. We find very good agreement with MaB{\textmu}lS-2 predictions. From the experiments visualised in Figs~\ref{fig:timescales_combined}-\ref{fig:lon_lat_diff}, we conclude that the results of our hierarchical inference almost always agree with MaB{\textmu}lS-2 predictions at a 1$\sigma$ level - both for individual $t_{\rm E}$ bin heights and for the estimated mean $\log t_{\rm E}$ values.

At the same time, we note that the $1\sigma$ confidence intervals for this estimation are large compared to differences between pairs of spatial bins, and hence these differences are not statistically significant within our inference (for neither individual $t_{\rm E}$ bin heights nor the estimated mean $\log t_{\rm E}$ values). We note the very strong dependence of $1\sigma$ interval widths on event count. We conclude that more observational data, both in terms of a higher event count and higher number of datapoints per event, is needed in order to increase the precision of $t_{\rm E}$ distribution inference.

We now use the spatially resolved timescale distributions from the MaB{\textmu}lS-2 simulation, created for this work, to analyse differences between fields in the central Milky Way regions and propose optimal observing strategies to capture their asymmetry.

\begin{figure}
	\includegraphics[clip, trim=0.35cm 0.2cm 0cm 0cm, width=1.02\columnwidth]{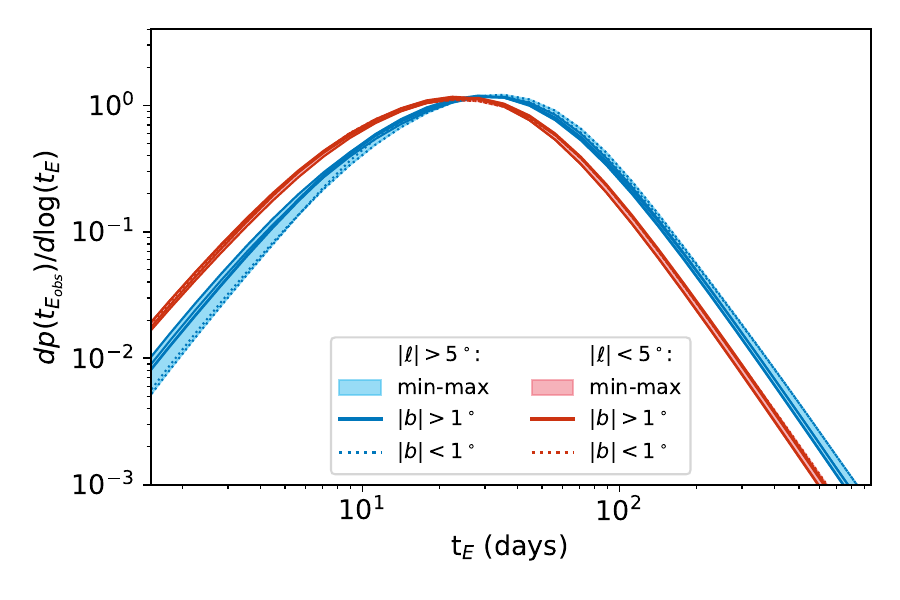}
  \caption{MaB{\textmu}lS-2 timescale distribution predictions for all 12 spatial bins. Red: central longitudes ($|\ell| < 5^\circ$), blue: non-central longitudes ($|\ell| > 5^\circ$). Low-alpha bands indicate min-max values. Individual distributions are plotted with dotted lines for $|b| <1^\circ$ and with solid lines for $|b| >1^\circ$.}
  \label{fig:mabuls_visualised}
\end{figure}

We first identify the optimal fields to capture significant difference between $t_{\rm E}$ distributions. In Fig~\ref{fig:mabuls_visualised}, we compare the MaB{\textmu}lS-2 predictions for all spatial bins. We note that the most striking feature is the splitting of the timescale distributions into two groups: the central ($|\ell| < 5^\circ$) and non-central ($|\ell| > 5^\circ$) longitudes, with the latter tending towards larger timescales. While all distributions for the central bins are very similar, careful analysis of the lower $t_{\rm E}$ edge would allow to identify difference between the low $|\ell|$ (dotted in Fig~\ref{fig:mabuls_visualised}) and high $|\ell|$ non-central regions (dotted in Fig~\ref{fig:mabuls_visualised}) -- that would require a very high-cadence (<1 day) survey. We note that among pairs of neighbouring bins, the highest difference in \edit{$\hat{t}_{\rm E}$} is between bins 7 and 8 \edit{($\Delta \hat{t}_{\rm E} \ = 9.1$ days)}. We recommend investigating the region covered by those bins ($|b|<1^\circ$, $0^\circ >$ \edit{$\ell$} $> -10^\circ$) to capture the \edit{spatial gradients in the $t_{\rm E}$ distribution} confidently. Among pairs of non-neighbouring bins, the highest differences in \edit{$\hat{t}_{\rm E}$} are between bins 3 and 8 \edit{($\Delta \hat{t}_{\rm E} \ = 9.5$ days)} and bins 2 and 8 \edit{($\Delta \hat{t}_{\rm E} \ = 9.4$ days)}.

We then identify the most optimal fields to capture asymmetry. This is useful in constraining the properties of asymmetric components of the Milky Way, especially the Galactic bar. While direct observations of stars get less complete and more confined to the highest absolute magnitudes with increasing distance, microlensing remains unconstrained by the brightness of the lens and is an ideal tool to study diverse stellar populations and structure in the central Galactic regions~\citep[e.g.,][]{Ki94,Ha00,Bi02,Ra07}.

We split the \edit{$\hat{t}_{\rm E}$} distribution from Fig~\ref{fig:spatial_binning} into a symmetric and asymmetric (in the respective coordinate) component: \edit{
\begin{align}
    &(\hat{t}_{\rm E})_{{\rm asym}, b}(\ell, b) = \nonumber \
    \hat{t}_{\rm E} (\ell, b) \ - (\hat{t}_{\rm E})_{{\rm sym}, b}(\ell, b)  = \\ &= \ \hat{t}_{\rm E} (\ell, b) - \frac{\hat{t}_{\rm E} (\ell, b) \ + \hat{t}_{\rm E}(\ell, -b)}{2} 
    \label{eq:asym_lat}
\end{align}}
and analogously, \edit{
\begin{equation}
    (\hat{t}_{\rm E})_{{\rm asym}, \ell}(\ell, b) = \hat{t}_{\rm E} (\ell, b) - \frac{\hat{t}_{\rm E} (\ell, b) \ + \hat{t}_{\rm E} (-\ell, b)}{2} \textnormal{.}
    \label{eq:asym_lon}
\end{equation}}

\begin{figure}
	\includegraphics[clip, trim=3.2cm 1.85cm 3.2cm 2cm, width=1.1\columnwidth]{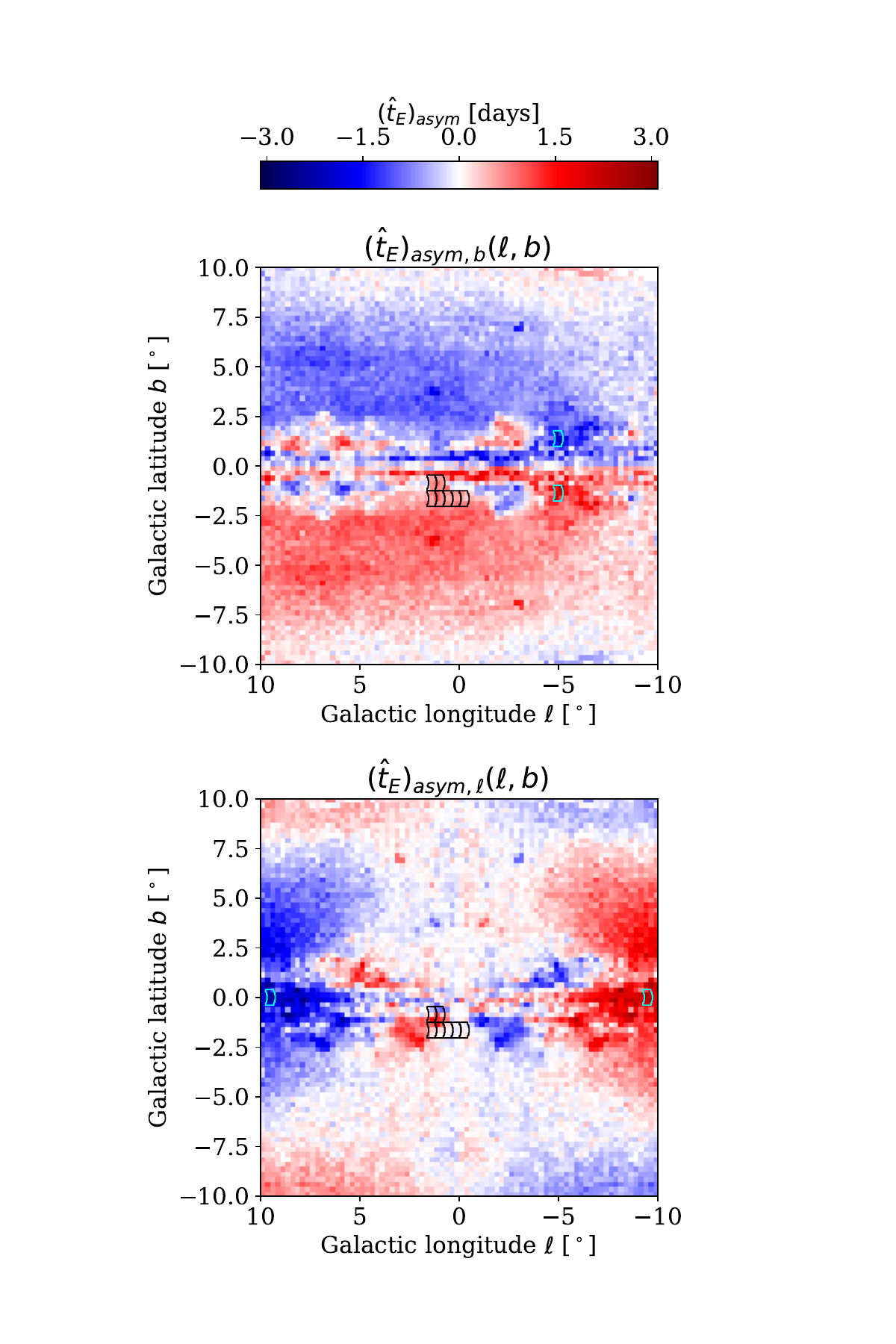}
  \caption{The asymmetric components of the spatial distribution of \edit{peaks of the lognormal $t_{\rm E}$ distribution} from MaB{\textmu}lS-2 predictions, as defined in Equations \ref{eq:asym_lat}-\ref{eq:asym_lon}. \\
  \textit{Upper:} Asymmetry in latitude \edit{ $(\hat{t}_{\rm E})_{{\rm asym}, b}(\ell, b)$}. \textit{Lower:} Asymmetry in longitude \edit{ $(\hat{t}_{\rm E})_{{\rm asym}, \ell}(\ell, b)$}. Black outlines: proposed fields of the \textit{Roman Space Telescope} Galactic Bulge Time Domain Survey. Cyan outlines: hypothetical \textit{Roman} fields chosen to maximise asymmetry. They are located at ($\ell = \pm 9,5^\circ$, \edit{$b=-0.125^\circ$}) and ($\ell = -5^\circ$, $b=\pm 1.375^\circ$) to capture asymmetry in longitude and latitude respectively.}
  \label{fig:asymmetry_lat_lon}
\end{figure}

We plot the asymmetric components in Fig~\ref{fig:asymmetry_lat_lon}. The \edit{$\hat{t}_{\rm E}$} asymmetry in latitude and longitude is at similar levels, with maximum absolute values of 
\edit{2.4} and  \edit{2.8} days and median absolute values of \edit{0.5} and \edit{0.3} days, respectively.

We propose that with a high-cadence near-infrared survey across fields with high \edit{$\hat{t}_{\rm E}$} differences, asymmetry could be captured at a statistically significant level. A suite of simulations similar to MaB{\textmu}lS-2 could be performed with varying Galactic bar properties (e.g. bar strength, bar angle), and those properties could be constrained by testing consistency with observed $t_{\rm E}$ distributions. This experiment is not yet possible with VVV data, but near-future near-infrared missions might provide sufficiently large and complete microlensing event databases.

The \textit{Nancy Grace Roman Space Telescope} is expected to be launched by May 2027. Its multi-band camera, WFI (Wide-Field Instrument), will observe in the near infrared, covering wavelengths up to 2300 nm\footnote{\url{https://roman.gsfc.nasa.gov/science/RRI/Roman_WFI_Reference_Information_20210125.pdf}} -- similarly to VVV's $K_s$ band. It is scheduled to carry out several surveys, of which the Galactic Bulge Time Domain Survey is of particular interest for microlensing. The survey will provide unprecedented cadence (of 15 minutes within 72-day seasons) and precision (astrometric: $\sim$ 1 mas, photometric: $\sim$ 10 milli-mag for a single observation -- increased when combining observations) in surveying the Galactic Bulge \citep[following][]{Penny2019, WFIRST2019, Gaudi2019}.

\citet{Ka22} have re-simulated one of the found VVV lensing events as seen by \textit{Roman}, finding $\sim$0.1\% precision in event timescale determination (compared to $\sim$10\% precision from VVV data for the same event), as well as <1\% precision in subsequent mass measurement. \citet{Sajadian2023} have also provided predictions of mass and timescales measurements of isolated stellar-mass black holes by \textit{Roman} up to $\leq$1\%. This level of precision will dramatically improve not only studies of single lensing objects, but also statistical analyses of the structure of the Galaxy.

As \textit{Roman} will observe with a very high cadence and reach magnitudes as faint as $\sim$ 23 for a single exposure\footnote{\url{https://roman.gsfc.nasa.gov/science/WFI_technical.html}}, the event yield is also expected to increase very significantly. Particularly, \citet{Penny2019} simulated the microlensing event yield of $\sim$ 27,000 events at impact parameter $|u|<1$ and $\sim$ 54,000 at $|u|<3$, distributed over 1.96 deg$^2$ of the field surveyed in \textit{Roman}'s Galactic Bulge Time Domain Survey (compared to the microlensing event yield of 1959 for the VVV search of \citet{Husseiniova2021}, and 1602 events distributed over 200 deg$^2$ used in this study). The high event yield, combined with high precision in $t_{\rm E}$ determination, will allow for carrying out a study similar to this, but with much narrower confidence intervals for the spatially resolved $t_{\rm E}$ distributions.

\edit{Efforts are underway to optimise the \textit{Roman} observing strategy even further, with e.g. \citet{Sajadian2023} demonstrating the expected wide timescale distribution of detected events, assuming additional sparse observations in the mid-survey gap. Additionally, strategies are proposed for filling the gaps in \textit{Roman} observations using other surveys, such as \textit{The Vera C. Rubin Observatory} Legacy Survey of Space and Time \citep{R2D2} or PRIME\footnote{\url{http://www-ir.ess.sci.osaka-u.ac.jp/prime/science/science.html}}.}

We overlay the proposed \textit{Roman} observing fields of the Galactic Bulge Time Domain Survey \citep{Penny2019}\footnote{using the code available at \url{https://github.com/mtpenny/wfirst-ml-figures}} on the asymmetry maps in Fig~\ref{fig:asymmetry_lat_lon}. We note that the asymmetry and overall \edit{$\hat{t}_{\rm E}$} variation is very small over the proposed fields. We set out to determine fields for a hypothetical community survey to capture the strongest asymmetry signal. We choose 2 pairs of \textit{Roman} WFI fields, centred on ($\ell = \pm 9,5^\circ$, \edit{$b=-0.125^\circ$}) and ($\ell = -5^\circ$, $b=\pm 1.375^\circ$) to capture asymmetry in longitude and latitude, respectively. \edit{To select those fields, we approximate the \textit{Roman} fields by a 0.5$^\circ$ x 0.75$^\circ$ rectangle, i.e. 2x3 MaB{\textmu}lS-2 pixels. Per each simulated field, we bin the respective 6 MaB{\textmu}lS-2 pixels together (weighted by event count) and fit $\hat{t}_{\rm E}$ to the resulting distribution. We iterate over the extent covered by MaB{\textmu}lS-2 in 0.25$^\circ$ intervals and we choose fields yielding the largest asymmetry.} We also overlay those fields on the maps in Fig~\ref{fig:asymmetry_lat_lon}. \newedit{We plan to further explore an optimal observing strategy, taking into account the characteristics of the \textit{Roman Space Telescope} and its surveys, in future work.}

In summary, we have developed a rigorous framework to analyse spatially resolved microlensing timescale distributions. This is a particularly promising avenue in studies of Galactic structure. Although the VVV survey has insufficient cadence to constrain properties of the Galactic bar, we expect forthcoming opportunities with {\it Roman} to make this eminently feasible.

\section*{Acknowledgments}
ZK thanks EC H2020 ORP grant no 101004719 for travel support and the Cambridge Streams group for lively discussions. ZK is a Fellow of the International Max Planck Research School for Astronomy and Cosmic Physics at the University of Heidelberg (IMPRS-HD). This work was performed under the auspices of the U.S. Department of Energy by Lawrence Livermore National Laboratory under Contract DE-AC52-07NA27344. The document number is LLNL-JRNL-854902. \edit{This work was supported by the LLNL-LDRD Program under Project 22-ERD-037. We thank the anonymous referee for valuable comments that improved the quality of this work.}

\section*{Data Availability}
The data underlying this article will be shared on reasonable request to the corresponding author.



\bibliographystyle{mnras}
\bibliography{refs} 




\appendix

\section{}
\label{sec:non-corr}

In this section, we discuss the excess bin heights occurring in the outer bins. We show how the convex shape of resulting timescale distributions originates at the stage of efficiency correction; we then show how those distributions dominated by the outer wings give unreliable $\langle t_{\rm E} \rangle$ estimates, which motivates our choice of fitting the peak of the timescale distribution instead.

\begin{figure*}
	\includegraphics[width=2.1\columnwidth]{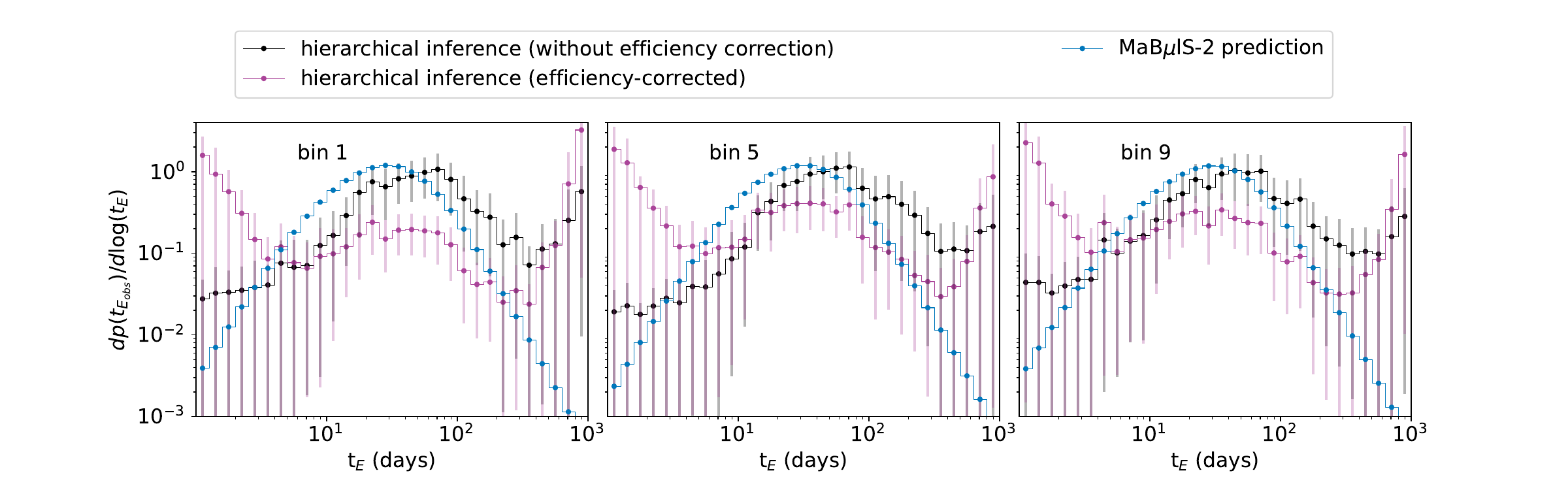}
  \caption{Timescale distributions from the hierarchical $\tE$ inference with (purple) and without (black) the efficiency corrections. All $t_{\rm E}$ distributions are normalised to 1. The errorbars indicate the 68\% confidence intervals on the $\tE$ bin heights. Respective MaB{\textmu}lS-2 predictions (blue) overplotted for comparison. \newline \textit{Left:} Timescale distributions for spatial fine bin 1. \textit{Centre:} Timescale distributions for spatial fine bin 5. \textit{Right:} Timescale distributions for spatial fine bin 9.}
   \label{fig:timescale_correction}
\end{figure*}

In Fig~\ref{fig:timescale_correction} we present the timescale distributions for 3 example bins before and after the efficiency correction. We note the efficiency correction heavily upweighs the outer wings of the $t_{\rm E}$ distribution, making the distribution convex.

\begin{figure*}
	\includegraphics[clip, trim=0cm 0cm 0cm 0cm, width=2.1\columnwidth]{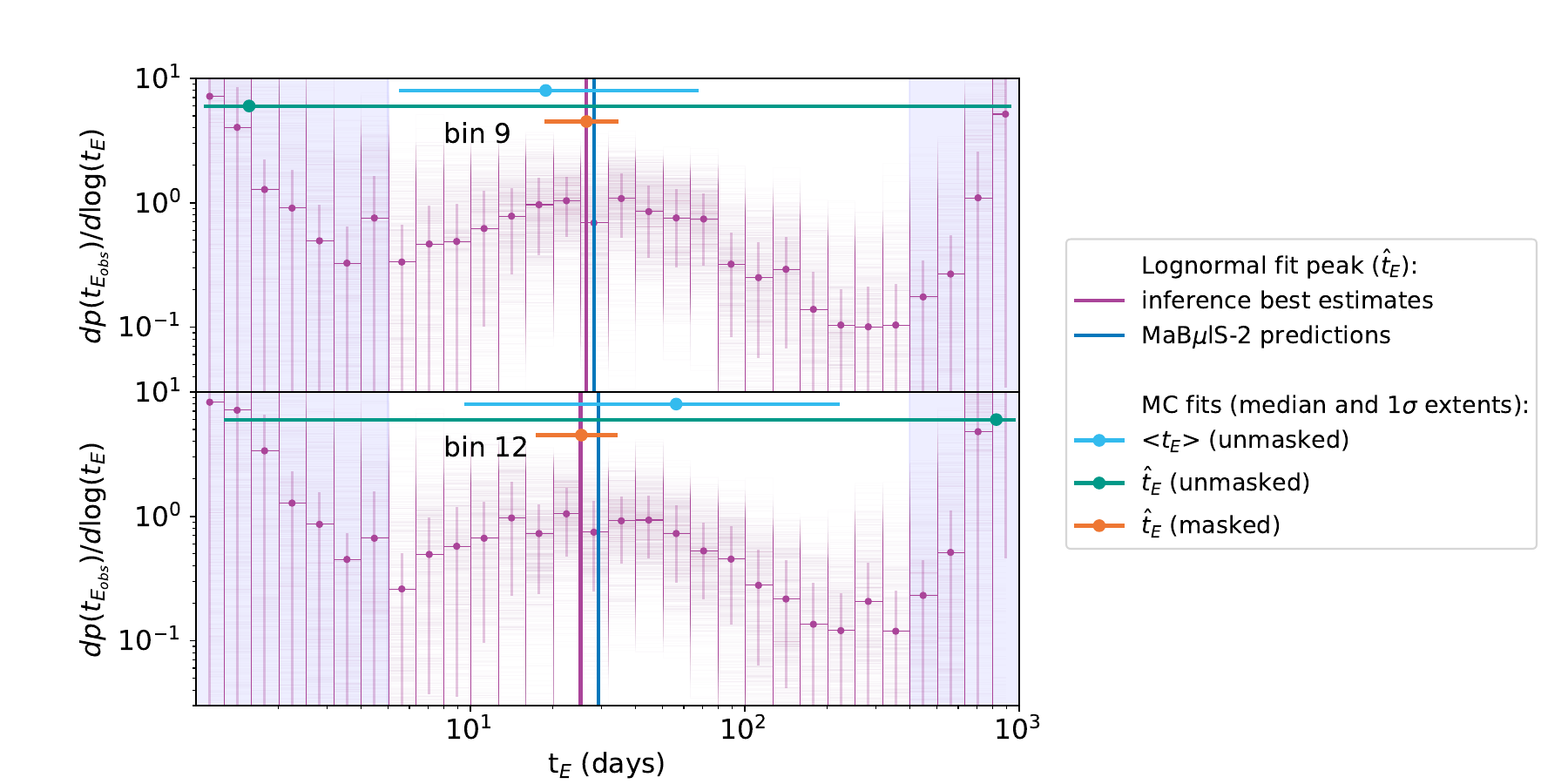}
  \caption{\edit{Comparison of different metrics to represent a typical event timescale and their sensitivity to noise for two example bins. Cyan: $\langle t_{\rm E} \rangle$ -- average over the full extent. Teal: mean of the lognormal distribution $\hat{t}_{\rm E}$ fitted to full extent. Orange: mean of the fitted lognormal distribution $\hat{t}_{\rm E}$ fitted to the unmasked extent.}
  Event timescale distribution, mask and $\hat{t}_{\rm E}$ values for the MaB{\textmu}lS-2 predictions and hierarchical inference results as in Fig~\ref{fig:timescales_combined}. Light purple represents 1000 randomly chosen samples $y_i(\log t_{\rm E})$. \edit{(Samples were re-drawn in case of runtime error (fit failed in 5000 iterations); this happened for 39.8\% (bin 9) and 20.5\% (bin 12) for the unmasked $\hat{t}_{\rm E}$ fits, and for 1.2\% and 1.9\% for the masked $\hat{t}_{\rm E}$ fits, respectively, and can result in underestimated errorbars especially for the unmasked $\hat{t}_{\rm E}$ part.)}}
   \label{fig:te_peak_mean}
\end{figure*}

In Fig~\ref{fig:te_peak_mean} we compare \edit{different metrics} with their respective confidence intervals by drawing 10000 random samples representing the histogram $y_i(\log t_{\rm E})$. We show results for example bins 9 and 12; we note that while the bins are symmetrical and their timescale estimates should be similar, the $\langle t_{\rm E} \rangle$ \edit{and unmasked $\hat{t}_{\rm E}$} estimates are very poorly constrained and dominated by the outer wings of the distribution, motivating our \edit{final choice of metric.}

\bsp	
\label{lastpage}
\end{document}